\documentclass[aps,prl,twocolumn,showpacs,superscriptaddress]{revtex4}
\usepackage{graphicx}

\begin{document}

\title{Ballistic thermal rectification}
     \author{Lifa~Zhang}
     \affiliation{Department of Physics and Centre for Computational Science and Engineering,
     National University of Singapore, Singapore 117542, Republic of Singapore }
      \author{Jian-Sheng~Wang}
     \affiliation{Department of Physics and Centre for Computational Science and Engineering,
     National University of Singapore, Singapore 117542, Republic of Singapore }
     \author{Baowen~Li}
    \altaffiliation{Electronic address: phylibw@nus.edu.sg}
     \affiliation{Department of Physics and Centre for Computational Science and Engineering,
     National University of Singapore, Singapore 117542, Republic of Singapore }
    \affiliation{NUS Graduate School for Integrative Sciences and Engineering,
      Singapore 117456, Republic of Singapore}
\date{8 Feb 2010}
\begin{abstract}
{We study ballistic thermal transport in three-terminal atomic nano-junctions
by the nonequilibrium Green's function method.  We find that there is
ballistic thermal rectification in asymmetric three-terminal structures
because of the incoherent phonon scattering from the control terminal. With
spin-phonon interaction, we also find the ballistic thermal rectification
even in symmetric three-terminal paramagnetic structures.}
\end{abstract}

\maketitle

{\emph{Introduction}} Phononics, the study of information processing
 and controlling of heat flow by phonons, is an emerging new field that
is attracting increasing attention \cite{Wang2008}. Specifically,
researchers have recently modeled and built thermal rectifiers
\cite{rectifiers}, thermal transistors \cite{transistor}, thermal
logical gates \cite{logicgate}, and thermal memory \cite{memory},
which are the basic components of functional thermal devices.  The
most fundamental phononic component is the thermal rectifier -- a
device that allows larger conduction in one direction than in the
opposite direction when it is driven far enough from equilibrium.
The effect of thermal rectification is well known to be realized by
combining the system inherent anharmonicity with structural
asymmetry \cite{rectifiers,transistor,logicgate,memory,segal2005,wu2009,ruokola2009,zhang20091}. Whether the rectification can happen in harmonic systems, that is, ballistic rectification, is
still unknown, although one recent paper \cite{hopkins2009}
discussed it, in which the authors get different heat conductances
by exchanging the heat baths with different structures connected to
an asymmetric center part, thus the different conductances come from
the totally different systems.  As we know, the thermal transport in
nanoscale materials, which is very promising for thermal devices,
can be regarded as ballistic because of their small sizes in
comparison with the phonon mean free path. Therefore, it is highly
desirable to investigate whether the ballistic thermal rectification
can be realized in harmonic systems, and to explore the necessary
conditions for thermal rectification.

The ballistic thermal transport in two-terminal junctions can be
described by the Landauer formula. Since the temperatures enter only
through the Bose distribution, it is obvious that if we reverse the
heat bath temperatures, the heat flux only changes sign, and no
rectification is expected. How about the ballistic thermal transport
in multiple-terminal junctions?  The theory for multiple-terminal
electric transport was proposed as the Landauer-B\"uttiker
conductance formula \cite{datta1995,buttiker,blanter2000}, and was applied to
thermal transport recently \cite{sun2002,ming2007,zhang20092}. From the electronic transport in three-terminal system \cite{blanter2000}, we know that the third terminal can introduce incoherence or phase breaking to the transport. So it
is our interest to investigate whether a multiple-terminal junction
is a proper option for ballistic thermal transport, that is, whether the incoherence through the third terminal can induce rectification effect. We will take the
nonequilibrium Green's function (NEGF) approach \cite{negf,refnegf,ruokola2009,ojanen2008}.
NEGF is widely applied to electronic and thermal transport, and is
successful to study the spin Hall effect and phonon Hall effect in
junctions \cite{Sheng2005,zhang20092}.

\begin{figure}
\includegraphics[width=2.8 in, height=1.4 in,,angle=0]{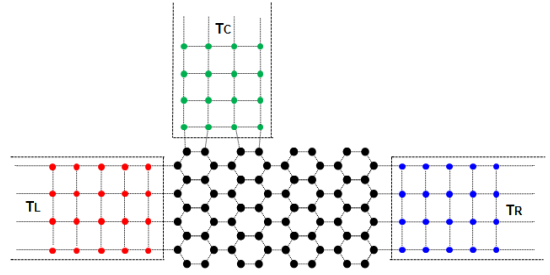}%
\caption{\label{fig1model}(Color online) The three-terminal junction setup to
study the ballistic thermal transport. The left and right leads have
temperatures $T_L$ and $T_R$, the control terminal lead is adjusted to be
$T_C$ or $T^{'}_C$ so that the heat flux from this lead is zero in the forward
($T_L=T_+$, $T_R=T_-$) or backward ($T_L=T_-$, $T_R=T_+$) process. $T_+$
and $T_-$ are the temperatures of the hot and cold baths, respectively.}
\end{figure}

{\emph{Model and Method}} We consider the ballistic thermal transport in a
three-terminal nano-junction as shown in Fig.~\ref{fig1model}, where a
two-dimensional lattice sample, which is a honeycomb lattice, is connected with three ideal
semi-infinite leads. The masses are coupled through nearest neighbors
by elastic springs (with longitudinal and transverse force constants).
We denote the center lattice as $N_R\times N_C$, $N_R,\,N_C$
correspond to the number of rows and columns, respectively. The external magnetic field can
be perpendicularly applied to this part. We use $N_{CL}$ to denote the number
of columns of the control lead and $N_{CD}$ to denote the number of columns
deviate from the middle of the center part; if $N_{CD}=0$, the whole setup is
symmetric. In Fig.~\ref{fig1model}, $N_R=9,\,N_C=8,\,N_{CL}=4,\,N_{CD}=-2$.
The Landauer-B\"uttiker formula can be expressed as,
\begin{equation}
J_\alpha   = \sum\limits_{\beta  \ne \alpha } {\int_0^\infty  {\frac{{d\omega }}
{{2\pi }}} \hbar \omega\, \tau _{\beta \alpha } (\omega )\bigl[n(T_\alpha  ) - n(T_\beta  )\bigr]}.
\end{equation}
Here, $\tau _{\beta \alpha }$ is the transmission coefficient from
the $\alpha$th bath to the $\beta$th bath; and
$n(T_\alpha)=(e^{\hbar\omega/k_B T_\alpha}-1)^{-1}$ is the Bose
distribution with $T_\alpha$ being the temperature of the $\alpha$th
heat bath. We set $\hbar=1$ and $k_B=1$ in the following
calculation.  Therefore, in the forward process, $T_L=T_+$,
$T_R=T_-$, we obtain
\begin{small}
\begin{equation}
J_+\!\!=\!\!\int\! {\frac{{\omega d\omega }}
{{2\pi }}\bigl\{ \tau _{RL} (\omega )[n(T_ +  ) - n(T_ -  )]}  + \tau _{CL} (\omega )[n(T_ +  ) - n(T_C )]\bigr\}
\end{equation}
\begin{equation}\label{jc}
J_C\!\!=\!\!\int\! {\frac{{\omega d\omega }}
{{2\pi }}\bigl\{ \tau _{LC} (\omega )[n(T_C ) - n(T_ +  )]}  + \tau _{RC} (\omega )[n(T_C ) - n(T_ -  )]\bigr\}
\end{equation}
\end{small}
\noindent Similarly we can obtain the heat fluxes $J_-$ and $J^{'}_C$ in the backward
process $T_L=T_-$, $T_R=T_+$. From the equations of $J_C=0$ and $J^{'}_C=0$ we
can obtain the temperatures of the control bath $T_C$ and $T^{'}_C$; inserting
them to the formulae of $J_+$ and $J_-$, by the definition of rectification as
\begin{equation}
 R =(J_+ - J_-)/ {\rm max }\{J_+, J_-\},
\end{equation}
we can calculate the rectification of this model. If the system is in the linear
response regime or in the classic limit, the heat flux from the $\alpha$th lead can be expressed as
$J_\alpha = \sum\limits_{\beta \ne \alpha } \sigma_{\beta \alpha} (T_\alpha -
T_\beta )$, we set $T_+=T_0+\Delta,\,
T_-=T_0-\Delta,\,T_C=T_0+\delta,\,T^{'}_C=T_0+\delta^{'}$, then we obtain
\begin{equation}
  \delta  =  - \delta ' = \frac{{\sigma _{LC}  - \sigma _{RC} }}
{{\sigma _{LC}  + \sigma _{RC} }}\Delta;
\end{equation}
\begin{equation}
  J_ +   = J_ -   = 2\Delta \left(\sigma _{RL}  + \frac{{\sigma _{CL} \sigma _{RC} }}
{{\sigma _{LC}  + \sigma _{RC} }} \right).
\end{equation}
So there is no rectification in the linear response regime or in the classic limit. In order to get
thermal rectification, we should consider the quantum regime out of linear response,
and the key work is to compute the transmission coefficients among the heat
baths.

In this paper, in addition to the structural asymmetry, we also can introduce the same spin-phonon interaction as in
Refs.~\cite{Sheng2006,Kagan2008,wang2009,zhang20092} in order to break
a time-reversal symmetry.
The Hamiltonian of our model can be written as
\begin{equation}\label{ham0}
H  = \sum\limits_{\alpha=0,L,R,C } H_{\alpha}
+ \sum\limits_{\beta=L,R,C}
{U_\beta ^T V_{\beta, 0} U_0 } +
  U_0^T AP_0,
\end{equation}
where $H_\alpha = \frac{1}{2}\left(P_\alpha ^T P_\alpha + U_\alpha ^T K_\alpha
U_\alpha\right)$.
The superscript $T$ denotes matrix transpose.
Here, $ 0, L, R, C$ correspond to the center region, left,
right, and control leads, respectively. $U_\alpha$ is a column vector for mass
reduced displacements in region $\alpha$,
$P_\alpha$ is the associated
conjugate momentum vector, and $K_\alpha$ is the force constant matrix, $V_{\beta,
0}=(V_{0, \beta})^T$ is the coupling matrix between the $\beta$th lead and the
central region.
$A$ is
a block diagonal matrix with diagonal elements $ \left(
{\begin{array}{*{20}c} 0 & h \\ -h & 0 \\
\end{array}} \right).$
$h$ is a model parameter which is supposed to be
proportional to the
magnetic field. In the
first part of this paper, we set $h=0$; it is a standard harmonic phononic system.

\begin{figure}[t]
\includegraphics[width=2.8 in, height=1.6 in,,angle=0]{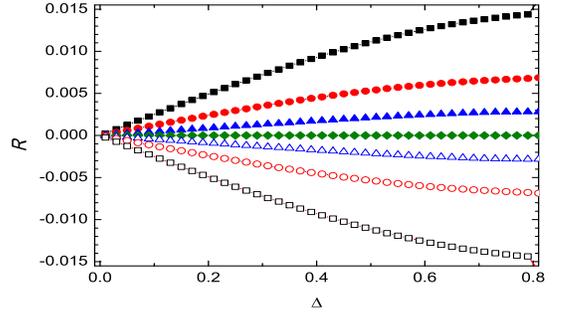}%
\caption{ \label{fig2rectncd} (Color online) Rectification as a function of
relative temperature difference of the hot and cold heat baths. The parameter
of the setup is $N_R=9,\,N_C=16,\, N_{CL}=2$. The temperature of the heat bath
are $T_+=T_0(1+\Delta)$ and $T_-=T_0(1-\Delta)$, where $T_0=0.2$ is the mean
temperature. The solid square, solid circle, solid triangle, diamond, hollow
triangle, hollow circle, hollow square correspond to $N_{CD}=-7$, -5, -3, 0,
3, 5 and 7, respectively.}
\end{figure}

The retarded Green's function for
the central region in frequency domain is \cite{zhang20092}
\begin{equation}
G^r [\omega ] = \Bigl[(\omega  + i\eta )^2  - K_0  - \Sigma^r [\omega ] -
A^2  + 2i\omega A\Bigr]^{ - 1}.
\end{equation}
Here, $\Sigma^r=\sum\limits_{\beta=L,C,R} {\Sigma_\beta ^r}$, and
$\Sigma_\beta ^r=V_{0, \beta} g_\beta ^r V_{\beta,0}$ is the self-energy due
to interaction with the heat baths, $ g_\beta ^
r=[(\omega+i\eta)^2-K_\beta]^{-1}$. $\eta$ is an infinitesimal real positive
quantity.
The surface Green's functions
of the leads $g_\beta^r$ are obtained following the algorithms of
Ref.~\cite{refnegf}.
 The transmission coefficient is
\begin{equation}\label{tran}
 \tau_{\beta \alpha } [\omega ] = {\rm{Tr}}(G^r \Gamma _\beta
G^a \Gamma _\alpha ),
\end{equation}
where, $ \Gamma _\alpha = i\bigl(\Sigma _\alpha ^r
[\omega ] - \Sigma _\alpha ^a [\omega ]\bigr)$, $G^a=(G^r)^{\dagger}$,
and $\Sigma _\alpha ^a=(\Sigma _\alpha ^r)^{\dagger}$.

{\emph{Results and Discussions}} Firstly, we consider the ballistic
thermal transport in an asymmetric structure without an external
magnetic field. In the following simulation, we set the longitudinal spring constant $k_L=1.0$, and the transverse one $k_T=0.25$. If the control lead is connected to the middle of upper edge of the center, that is, $N_{CD}=0$, the forward process and backward one are exactly the same; no rectification will be expected, as shown in Fig.~\ref{fig2rectncd} (the diamond symbols).  We will obtain the same result of no rectification if we replace the center honeycomb lattice with the square lattice same as the leads, when the whole system is symmetric wherever we put the control lead.  If the control lead
moves away from the center, the rectification effect appears. When the lead is moved the same distance to the left or right, the rectification coefficient has the same magnitude but opposite sign, which is because that the two cases only exchange the value of $J_+$ and $J_-$.  If the distance between the control lead and the middle of the center part is longer, the rectification effect is larger. In Fig.~\ref{fig2rectncd}, we can see that the case of $N_{CD}=\pm 7$, when the control lead is next to left or right lead, has biggest rectification.  The rectification increases with the temperature difference at far-from-linear-response regime.
\begin{figure}[t]
\includegraphics[width=2.8 in, height=1.6 in, angle=0]{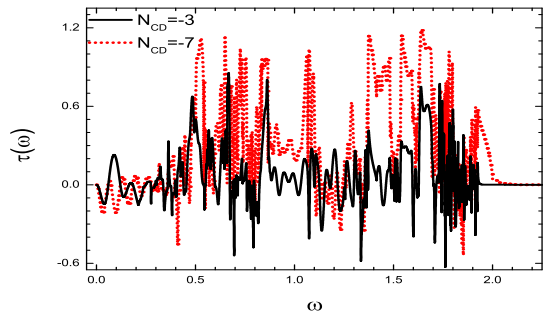}%
\caption{ \label{fig3trandif} (Color online) The difference of transmission coefficients: $\tau_{LC} - \tau_{RC}$, as a function of frequency. The parameter of the
setup is $N_R=9,\, N_C=16,\, N_{CL}=2$. The solid, dot curves correspond to
 $N_{CD}=-3$ and $N_{CD}=-7$, respectively. . }
\end{figure}

From the above formulae, we find if the transmissions $\tau_{LC}$
and $\tau_{RC}$ are linearly dependent (here
$\tau_{\alpha\beta}=\tau_{\beta\alpha}$ because of time reversal
symmetry), the rectification will be zero. Figure~\ref{fig3trandif}
shows that the transmission coefficients $\tau_{LC}$ and $\tau_{RC}$
are not linearly dependent, so that the rectification is nonzero. In
some frequency domains, $\tau_{LC}>\tau_{RC}$; and in other
frequency domains, $\tau_{LC}<\tau_{RC}$. With the distance
increasing, that is, from $N_{CD}=-3$ to $N_{CD}=-7$, the difference
of $\tau_{LC}$ and $\tau_{RC}$ enlarges in most part of the whole
frequency domain; so that the rectification coefficient increases.
Because of scattering from the third bath -- the control lead, the
total thermal transport from the left lead to the right one is
partially incoherent. This phonon incoherence induces rectification.
In the whole system the Hamiltonian is quadratic, that is, there is
no nonlinearity or anharmonicity, but we still can obtain
rectification.  Therefore, the phonon incoherence,  which can be
induced by either nonlinearity or scattering lead, is the
necessary condition for thermal rectification.
\begin{figure}[t]
\includegraphics[width=3.0 in, height=2.0 in, angle=0]{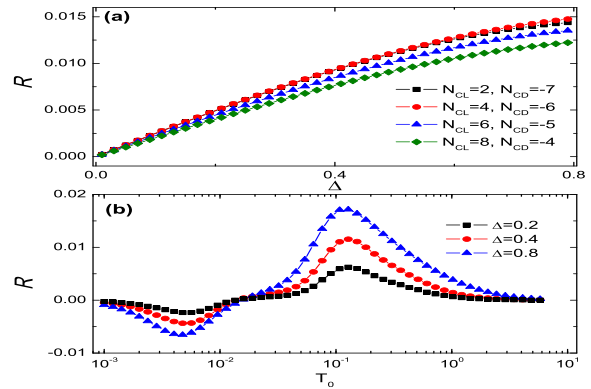}%
\caption{ \label{fig4rectnclT0} (Color online) (a) Thermal rectification as
function of relative temperature difference $\Delta$ for different width of
control lead, at $T_0=0.2$; (b) Thermal rectification as function of mean
temperature for different relative temperature
difference. $N_{CL}=2,\,N_{CD}=-7$. For both (a) and (b), $N_R=9, N_C=16$,
$T_+=T_0(1+\Delta)$ and $T_-=T_0(1-\Delta)$.  }
\end{figure}

The control lead acts as a scattering source, which makes the phonon
transport incoherent, so that the rectification comes out. However,
the width of the control lead does not quite affect the whole
thermal transport, which is shown in Fig.~\ref{fig4rectnclT0}(a). We
make the control lead next to the left lead, and find that the
rectification changes little when we increase the width of the lead.
If the width of control lead increases further, the rectification
decreases because the asymmetry decreases.
Figure~\ref{fig4rectnclT0}(b) shows the rectification dependence on
temperature, and reproduce the reversal of rectification found in
Ref.~\cite{zhang20091}. At a low temperature, the contribution to
thermal transport only comes from the low frequency phonons; if the
temperature increases, more high frequency phonons will contribute
to the heat transport. From Fig.~\ref{fig3trandif}, the relations
between transmissions $\tau_{LC}$ and $\tau_{RC}$ in low frequency
domain and high frequency domain are opposite, so that the
rectification reverses with the temperature increasing.
When the temperature increases further, the system will go to the classic limit, the rectification disappears.
\begin{figure}[b]
\includegraphics[width=2.8 in, height=2.0 in, angle=0]{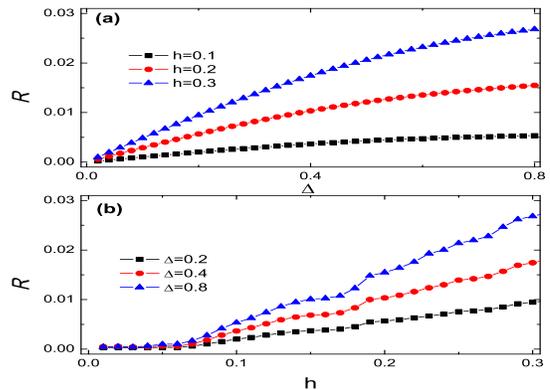}%
\caption{ \label{fig5magnrectdth} (Color online) (a) Thermal rectification as
function of relative temperature difference $\Delta$ for different external
magnetic fields. (b) Thermal rectification as a function of magnetic field
$h$.  For both (a) and (b): $N_R=9,\, N_C=16,\, N_{CL}=2,\,
N_{CD}=0$. $T_+=T_0(1+\Delta)$, $T_-=T_0(1-\Delta)$. $T_0$=0.2.  }
\end{figure}

\begin{figure}[t]
\includegraphics[width=2.8 in, height=1.6 in, angle=0]{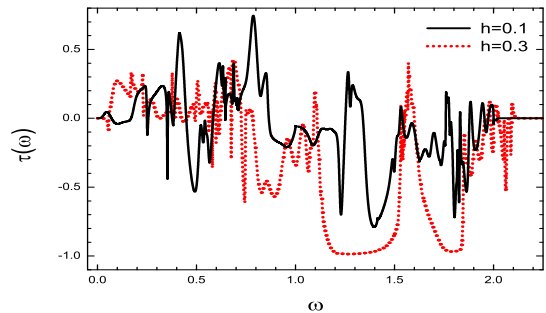}
\caption{ \label{fig6tranhdif} (Color online) The difference of transmission coefficients:
$\tau_{LC} - \tau_{RC}$, as a function of frequency for different applied magnetic fields. The parameter of the setup is $N_R=9,\,N_C=16,\, N_{CL}=2,\,
N_{CD}=0$.  The solid, dot curves correspond to $h=0.1$ and $h=0.3$, respectively. }
\end{figure}

From the previous work \cite{rectifiers,segal2005,zeng08} on thermal
rectification, we know that in order to get rectification, we need
the structural asymmetry. However, in the nanoscale rectifier, it is
not easy to control the structural asymmetry or not easily
distinguish the rectification direction by the structural asymmetry.
Is there any other means to introduce asymmetry to induce
rectification? From the study of phonon Hall effect
\cite{Sheng2006,Kagan2008,wang2009,zhang20092}, it is known that the
magnetic field can influence the thermal transport by the
spin-phonon interaction. Thus the magnetic field can break the
symmetry of the phonon transport. We apply an external magnetic
field perpendicular to the center part of a symmetric structure to
study the ballistic thermal transport, the results are shown in
Fig.~\ref{fig5magnrectdth}. The thermal rectification effect as a
function of the temperature difference is shown in
Fig.~\ref{fig5magnrectdth}(a). $R$ increases with the temperature
difference, and can be about $3\%$ if $\Delta=0.8$ and $h=0.3$ at
$T_0=0.2$. Figure~\ref{fig5magnrectdth}(b) shows that the
rectification can monotonically increase with the external magnetic
field in the range of $h=0\sim0.3$. The applied magnetic field
breaks the symmetry of the phonon transport system through the
spin-phonon interaction, so the transmission coefficient from the
control lead to left one is different from that to right one.
Figure~\ref{fig6tranhdif} shows that the two transmission coefficients
are not linearly dependent; and with the increasing of magnetic
field, the difference of these two transmissions will enlarge, which
induce bigger rectification effect.  Here all quantities are
dimensionless, if we use the parameters for real materials, the
rectification coefficient may be different, but would be likely in
the range of few per cents.

Although all our values for rectification are small, if we can devise a chain
of such ballistic rectifiers, the total rectification can be as large as we
need. Because for most nanoscale materials the thermal transport is
ballistic and temperature can be applied far from linear response
regime, our prediction can be tested by experiments. The asymmetric quasi-two-dimensional nano-structure of any material can have ballistic thermal rectification, such as a graphene sheet. Or the symmetric structure with spin-obit interaction, such as paramagnetic dielectrics, can also have ballistic rectification applied an perpendicular magnetic field.

{\emph{Conclusions}} Using the nonequilibrium Green's function method, we have studied ballistic thermal transport in three-terminal atomic nano-junctions. Adjusting
the temperature of the control lead in order to make the heat flux from this
lead be zero, we can calculate the thermal rectification effect. In the quantum regime 
out of linear response, there is ballistic thermal rectification in asymmetric
three-terminal structures because of incoherent phonon scattering from the
control terminal. Through spin-phonon interaction, we also find the ballistic
thermal rectification in symmetric three-terminal paramagnetic structures
applied an external magnetic field. Therefore, not the nonlinearity, but the
phonon incoherence, which can be induced by nonlinearity or scattering
boundary or scattering lead, is the necessary condition for thermal
rectification. Another necessary condition is asymmetry, not necessarily being
structural asymmetry, which can be introduced by an applied external magnetic
field through the spin-phonon interaction.

{\emph{Acknowledgements}} We thank Xiaoxi Ni, Jie Ren and Jie Chen for fruitful
discussions. LZ and BL are supported by the grant
R-144-000-203-112 from Ministry of Education of Republic of Singapore.
JSW acknowledges support from a faculty
research grant R-144-000-257-112 of NUS.

\end{document}